\documentclass{article}
\usepackage{graphicx}
\def\p{\partial}

\def\g{\gamma}

\def\ld{\lambda}
\def\Ld{\Lambda}

\def\s{\sigma}

\def\b{\beta}

\def\a{\alpha}

\def\pdellx'{\frac{\partial}{\partial x'}}
\def\pdellw'{\frac{\partial}{\partial w'}}

\newcommand{\be}{\begin{equation}}
\newcommand{\ee}{\end{equation}}
\def\bed{\begin{displaymath}}
\def\eed{\end{displaymath}}
\def\bea{\begin{eqnarray}}
\def\eea{\end{eqncrray}}
\def\[{$$}
\def\]{$$}

\begin{document}
\title{ YANG-MILLS GRAVITY IN FLAT SPACE-TIME, II.  \\
 GRAVITATIONAL RADIATIONS AND LEE-YANG FORCE FOR ACCELERATED 
 COSMIC EXPANSION }

\author{ JONG-PING HSU \\
Department of Physics,
 University of Massachusetts Dartmouth \\
 North Dartmouth, MA 02747-2300, USA\\
E-mail: jhsu@umassd.edu}


\maketitle
{\small  Within Yang-Mills gravity with translation group $T(4)$ in 
flat space-time, the invariant action involving quadratic 
translation gauge-curvature leads to quadrupole radiations which are 
shown to be 
consistent with experiments.  The radiation power turns out to be the same as that 
in Einstein's gravity  to the second-order approximation.  We also 
discuss an interesting physical reason for the accelerated cosmic expansion
based on the long-range Lee-Yang force 
of $U_{b}(1)$ gauge field 
associated with the established conservation law of baryon number.  
We show that the Lee-Yang force can be 
related to a linear potential $\propto r$, provided the  gauge field 
satisfies a fourth-order differential 
equation.  Furthermore, we consider an experimental test of the Lee-Yang force 
related to the accelerated cosmic expansion.  The necessity
of generalizing Lorentz transformations for 
accelerated frames of reference
and accelerated Wu-Doppler effects are briefly discussed.}

\bigskip
 

\bigskip


 
\section{ Introduction}

Right after the creation of the Yang-Mills theory in 1954, Utiyama 
immediately generalized the gauge field with SU(2) group to a general symmetry 
group with N generators and proposed the gauge-invariant 
 interpretation of all interactions.\cite{1,2}  In his pioneer work, Utiyama paved 
 the  way for far-reaching research on gauge theories of gravity and 
quantum gravity.  In such gauge theories of gravity, 
Utiyama and others followed the usual approach of general 
relativity to formulate their theories within the framework 
of curved spacetime.  However, the quantum aspect of general 
 relativity in curved spacetime encountered long-standing 
 difficulties because there is hardly any 
common ground between general relativity and quantum mechanics, as 
Wigner put it.\cite{3}  Dyson also stressed that the most glaring 
incompatibility of concepts in contemporary physics is that between 
the principle of general coordinate invariance and a 
quantum-mechanical description of all nature.\cite{4}  This incompatibility 
motivates the present investigation of Yang-Mills gravity with 
translation gauge symmetry within the framework of flat 
spacetime.  

Yang-Mills gravity is logically independent of the conventional 
theory of gravity (or general relativity).  The formulation of 
Yang-Mills gravity is based on bona fide local field theory in flat 
spacetime rather than  in curved spacetime.  Thus Einstein's field 
equation and Riemannian (or Robertson-Walker) metric are not 
postulate and, hence, cannot be used in the prersent formulation of gravity.

The theory follows a close analogy with the 
gauge invariant electrodynamics.   
Namely, the relevant symmetry leads, via Noether theorem, to a 
conserved quantity 
which is precisely the source for generating the field:
\bigskip

U(1) symmetry $\to$ conserved electric charge $\to$ the source of the 
electromagnetic potential field $A_{\mu}$. 

[with the replacement $\p_\mu \to \p_\mu - ieA_\mu$].  
\bigskip

T(4) spacetime translation symmetry $\to$ conserved 
energy-momentum tensor $\to$ the source of the gravitational field 
$\phi_{\mu\nu}$.  

[with the replacement $\p_\mu
\to \p_\mu  - ig\phi_{\mu}^{\nu}(i\p_\nu)$, \ \  $i\p_\nu$= generators 
of T(4)].\cite{5} 
\bigskip
 
In a recent work (paper I),\cite{5} I follow Yang-Mills' approach for 
internal gauge groups to 
discuss a generalized gauge theory for the `external' spacetime 
translation group, whose generators do not have constant matrix 
representations, in sharp contrast to the internal groups.  The 
above analogy between the internal group U(1) and the external 
spacetime group T(4) reveals two fundamental 
differences regarding electromagnetic and gravitational 
forces: 

(i) Although the electromagnetic coupling constant $e$ is 
dimensionless ($c=\hbar=1$), 
the basic gravitational coupling constant $g$ must have the dimension of 
length. 

(ii) While the electromagnetic force involves both 
attractive and repulsive forces, the gravitational force can have 
only one of them.\cite{5} 
\noindent

The present 
formulation of gauge theory with a translation gauge symmetry differs 
from all previous formulations with translation symmetry\cite{6,7} because (a) 
it is formulated within the framework of flat spacetime, and (b) it 
involves the Yang-Mills-type Lagrangian with quadratic gauge 
curvature, so that the interaction vertices of gravitons in Feynman 
diagrams are much simpler than those in general relativity.  Thus,
the present theory of Yang-Mills gravity has 
a common ground with quantum mechanics and quantum theory of gauge 
fields, and can be quantized just as usual fields.  It is 
expected that the T(4) gauge symmetry in flat spacetime will help to 
reduce the degrees of divergence in higher-order Feynman diagrams of 
Yang-Mills gravity and, hence, has a better high-energy behavior than 
that of general relativity.  I have shown that 
the gauge-invariant action 
with quadratic gauge-curvature in flat space-time can produce 
good agreement with classical experiments such as the perihelion 
shift of Mercury, bending of light and so on.\cite{5}  In this paper, I present
the calculation of
gravitational quadrupole radiation and show its 
consistency with observations in section 2.  Furthermore, to be consistent with the recent 
discovery of the 
accelerated cosmic expansion, I also consider the baryonic $U_{b}(1)$ gauge field 
which can produce a repulsive long-range force between baryon matter 
in flat spacetime.  The gauge group for the whole theory is 
$T(4)\times U_{b}(1)$.  Such a long-range
`Lee-Yang force' can be the physical reason for the   
accelerated cosmic expansion, because there is a modified $U_{b}(1)$ gauge 
invariant Lagrangian which leads to a linear cosmic potential $\propto r$. 
 This is discussed in section 4.  In section 5, we discuss an
 experimental test of the cosmic Lee-Yang
 force based on the accelerated Wu-Doppler effect.

\section{\large Gravitational Quadrupole Radiations}
\noindent

Yang-Mills gravity is formulated for general frames of reference 
(i.e., inertial and non-inertial frames).\cite{5}  For simplicity, we choose 
 inertial frames (in which 
$P_{\mu\nu} = \eta_{\mu\nu}=(+,-,-,-)$ and $D_{\mu}=\p_{\mu}$) to discuss the 
gravitational quadrupole radiation.  Ordinary 
matter contains baryons such as protons and 
neutrons (or up- and down-quarks).   Let us 
consider the $T(4)\times U_{b}(1)$ gauge invariant action $S=\int L d^{4}x$. 
The Lagrangian $L$ involves  the gravitational tensor field $\phi_{\mu\nu}$, 
a (baryonic) fermion field $\psi$ and  the $U_{b}(1)$ gauge  field 
associated with the conserved baryon numbers, \cite{5}  
$$L= \frac{1}{2g^2}\left (C_{\mu\a\b}C^{\mu\b\a}- 
 C_{\mu\a}^{ \ \ \  \a}C^{\mu\b}_{ \ \ \  \b} \right) $$ 
\be
+\frac{i}{2}\left [~ \overline{\psi} \g^{\mu} (\Delta_{\mu} 
\psi-ig_{b}B_{\mu}\psi) - 
(\Delta_{\mu} \overline{\psi}+ig_{b}B_{\mu}
\overline{\psi}) \g^{\mu} \psi \right] -
m\overline{\psi}\psi +L_{B},
\ee
\be
L_{B}= -\frac{L_s^2}{4} \Delta_{\ld} B_{\mu\nu} \Delta^\ld B^{\mu\nu}, 
  \ \ \ \ \ \    B_{\mu\nu}=\Delta_{\mu} B_{\nu} -\Delta_{\nu}B_{\mu},
\ee
$$ C^{\mu\nu\a} = J^{\mu\ld}(\p_\ld 
J^{\nu\a}) - J^{\nu\ld}(\p_\ld J^{\mu\a}),  \ \ \ \ \ 
C_{\mu\a\b}C^{\mu\b\a}=(1/2) C_{\mu\a\b}C^{\mu\a\b},$$
$$\Delta_\mu \psi = J_{\mu\nu}\p^\nu \psi, \ \ \ \ \ \  
J_{\mu\nu}=\eta_{\mu\nu}+ 
g \phi_{\mu\nu} = J_{\nu\mu} , \ \ \ \   c= \hbar = 1.$$
 
For the gravitational quadrupole radiation, we consider only the 
tensor field $\phi_{\mu\nu}$ and ignore the 
$U_{b}(1)$ gauge field $B_{\mu}$ because its coupling to baryon 
matter is much more weaker than that of gravitation.  Moreover, it suffices to
calculate the gravitation radiations  
to the second order in $g\phi_{\mu\nu}$.  As usual,
we impose the gauge condition
\be
\p^\mu \phi_{\mu\nu} = \frac{1}{2} \p_\nu \phi ^\ld_\ld.
\ee
The gauge invariant action with the 
Lagrangian (1) leads to the gravitational tensor field equation in inertial 
frames,\cite{5}
\be
H^{\mu\nu} =   g^2 T^{\mu\nu},
\ee
$$H^{\mu\nu} \equiv \p_\ld (J^{\ld}_\rho C^{\rho\mu\nu} - J^\ld_\a 
C^{\a\b}_{ \ \ \ \b}\eta^{\mu\nu} + C^{\mu\b}_{ \ \ \ \b} J^{\nu\ld})    $$
$$- C^{\mu\a\b}\p^\nu J_{\a\b} + C^{\mu\b}_{ \ \ \ \b} \p^\nu J^\a_\a -
 C^{\ld \b}_{ \ \ \ \b}\p^\nu J^\mu _\ld,$$
where $\mu$ and $\nu$ should be made symmetric and we have used the 
identities
\be
C_{\mu\a\b}= -C_{\a\mu\b}, \  \  \  \  \  \   C_{\mu\a\b} +  C_{\a\b\mu} +  C_{\b\mu\a} = 0.
\ee
It is not necessary 
to write this symmetry of $\mu$ and $\nu$ in (4) explicitly  for 
the following discussions of gravitational radiations.
The energy-momentum tensor $T^{\mu\nu}$ in equation (4) is given by
\be
T^{\mu\nu} = \frac{1}{2}\left[ \overline{\psi} i\gamma^\mu \p^\nu \psi -
i(\p^\nu \overline{\psi}) \gamma^\mu \psi \right].  
\ee
 
For weak fields in inertial frames with 
the gauge condition (3), the field equation can be linearized as follows:
\be
\p_\ld \p^\ld \phi^{\mu\nu} - \p^\mu \p_\ld \phi^{\ld\nu} + 
 \p^\mu \p^\nu \phi^\ld_\ld - \p^\nu \p_\ld \phi^{\ld\mu} =  g (T^{\mu\nu} 
- \frac{1}{2}\eta^{\mu\nu} 
 T^\ld_\ld), 
\ee
where we have used  
$J^{\mu\nu} = \eta^{\mu\nu} + g \phi^{\mu\nu}.$   
With the 
help of the gauge condition (3), (7) can be written as
\be
\p_\ld   \p^\ld \phi_{\mu\nu} = g (T_{\mu\nu}  - \frac{1}{2} 
\eta_{\mu\nu} T^\ld_\ld ) \equiv g S_{\mu\nu}, \ \ \ \ \ \  
g=\sqrt{8\pi G_{N}},
\ee
where $G_{N}$ is the Newtonian 
gravitational constant.  To the first-order approximation, we obtain\cite{5}
\be
g\phi_{00} = g\phi_{11} = \frac{G_{N} m}{r},  \ \ etc. 
\ee 
As usual, the energy-momentum 
tensor $T^{\mu\nu}$ is independent of $\phi^{\mu\nu}$ and  satisfies   the conservation law,
\be
\p_{\mu} T^{\mu\nu} = 0,
\ee
in the weak field approximation.

From equation (8), one has the usual retarded potential
\be
\phi_{\mu\nu} ({\bf x},t) = \frac{g}{4 \pi} \int d^3 x' 
\frac{S_{\mu\nu}({\bf x}', t-|{\bf x} - {\bf x}'|)}{|{\bf x} - {\bf 
x}'|}, 
\ee
$$   {\bf x} \equiv {\bf r}, \ \ \  {\bf x}' \equiv {\bf r}', 
 \ \ \ x^{\mu}=(w,x,y,z),  \ \ \  w=ct=t,  $$
which is generated by the source $S_{\mu\nu}$ in (8).   This equation is 
usually used to discuss the 
gravitational radiation.  When one discusses the 
radiation in   the wave zone at a distance much larger  than the 
dimension of the source, the solution can be approximated by a 
plane wave,\cite{8}
\be
\phi_{\mu\nu} (x) = e_{\mu\nu} exp(-ik_\ld x^\ld) +  e^*_{\mu\nu} exp(ik_\ld 
x^\ld),
\ee
where $e_{\mu\nu}$ is the polarization tensor.  The plane wave property and the usual gauge 
condition (3) lead to
\be
k_\mu k^\mu = 0, \ \ \ \ \ \   k_\mu e^\mu_\nu = \frac{1}{2} k_\nu 
e^\mu_\mu, \ \ \ \ \ \     k^\mu = \eta^{\mu\nu} k_\nu.
\ee
For the symmetric polarization tensor, $e_{\mu\nu}=e_{\nu\mu}$,
 of  the massless tensor field in flat spacetime, there 
are only two physical states with 
helicity $\pm 2$ which are invariant under the Lorentz transformation.

Let us write $T_{\mu\nu}({\bf x},t)$ in terms of a Fourier 
integral,~\cite{8}
\be
T_{\mu\nu}({\bf x},t) = \int_{0}^{\infty}T_{\mu\nu}({\bf 
x},\omega)e^{-i\omega t} d\omega + c.c.
\ee
The retarded field emitted by a single Fourier     component 
$T_{\mu\nu}({\bf x},t) = [T_{\mu\nu}({\bf x},\omega)e^{-i\omega t} + c.c.]$ 
is given by
\be
\phi_{\mu\nu}({\bf x},t) = \frac{g}{4\pi} \int\frac{d^3x'}{{\bf  x}-{\bf x}'} 
S_{\mu\nu}({\bf x}',\omega) exp(-i\omega t + i\omega |{\bf x} - {\bf x}'|)+ 
c.c.
\ee
$$S_{\mu\nu}({\bf x},\omega) = T_{\mu\nu}({\bf x},\omega) 
 - \frac{1}{2}\eta_{\mu\nu}T({\bf x},\omega), \ \ \   T = 
 T^\ld{_\ld}.$$
 
The energy-momentum tensor $t_{\mu\nu}$ of gravitation is defined by the 
exact field equation (4) written in the following form,
\be
\p^\ld \p_{\ld} 
\phi^{\mu\nu}= g(T^{\mu\nu} - t^{\mu\nu}). 
\ee
Thus we have 
\be
t^{\mu\nu}=\frac{1}{g^2}[C^{\rho\mu\nu}\p_{\ld}J^{\ld}_\rho + \p_\rho(g 
\phi^{\rho\ld}\p_\ld J^{\mu\nu}) + g\phi^{\ld}_\rho \p_\ld 
(J^{\rho\a} \p_\a J^{\mu\nu})
\ee
$$- J^\ld{_\rho} \p_\ld (J^{\mu\a} \p_\a J^{\rho\nu})
- C^{\mu\b\a} \p^\nu J_{\a\b} - \eta^{\mu\nu} \p_\ld (J^\ld{_\rho} 
C^{\rho\b}_{ \ \ \  \b}) + \p_\ld (C^{\mu\b}_{ \ \ \  \b} J^{\nu\ld})
$$
$$
+ C^{\mu\b}_{ \ \ \  \b} \p^\nu J^\ld{_\ld} - C^{\ld\b}_{ \ \ \  \b} 
\p^\nu J^\mu _{\ld}]. $$
To a second order approximation, we obtain the energy-momentum tensor of 
the gravitational field,
\be
t^{\mu\nu} = t^{\mu\nu}_1 + t^{\mu\nu}_2, 
\ee
\be
t_1^{\mu\nu}= (\p_\ld \phi) \p^\ld \phi^{\mu\nu} - \frac{1}{2}(\p_\ld \phi) 
\p^\mu \phi^{\ld\nu} + 2 \phi^{\ld\s} \p_\ld \p_\s \phi^{\mu\nu} - (\p_\ld 
\phi^{\mu\s}) \p_\s \phi^{\ld\nu}
\ee
$$-\phi^\ld_\rho \p_\ld \p ^\mu \phi^{\rho\nu} - \frac{1}{2} \phi^{\mu\s} 
\p_\s \p^\nu \phi - (\p^\mu \phi^{\b\a}) \p^\nu \phi_{\a\b} + (\p^\b 
\phi^{\mu\a}) \p^\nu \phi_{\a\b},$$
\be
t_2^{\mu\nu} = -\frac{3}{4}(\p_\ld \phi) \p^\ld \phi \eta^{\mu\nu} - 
\phi^{\ld\s} \p_\ld \p_\s \phi \eta^{\mu\nu} + (\p_\ld \phi^{\b\s}) \p_\s 
\phi^\ld{_\b} \eta^{\mu\nu}
\ee
$$
+ \frac{1}{2}(\p^\nu \phi^{\mu\ld}) \p_\ld \phi +\frac{1}{2} \phi^{\nu\ld} \p^\mu 
\p_\ld \phi + \phi^{\mu\s} \p_\s \p^\nu \phi $$
$$- (\p^\nu \phi^{\b\s}) \p_\s 
\phi^\mu_{\b} -\phi^{\b\s} \p _\s \p^\nu \phi^\mu_{\b} + \frac{3}{4} (\p^\nu \phi) \p^\mu \phi, $$
where $\phi \equiv \phi^{\ld}_{\ld}$, and we have used the gauge condition (3).  The energy-momentum tensor $t_1^{\mu\nu}$ and $t_2^{\mu\nu}$ are 
respectively contributed from the first and the second quadratic 
gauge-curvatures (i.e., $C_{\mu\a\b}C^{\mu\b\a}$ and 
$- C_{\mu\a}^{ \ \ \  \a}C^{\mu\b}_{ \ \ \  \b}$
respectively) in the Lagrangian (1).  One can use the plane wave 
solution (12) and the gauge condition (13) to calculate the energy-momentum 
tensors (18) in an inertial frame.   This 
complicated result can be simplified by taking the 
average of $t^{\mu\nu}$ over a region of space and time much larger than 
the wavelengths of the radiated waves.\cite{8}  After such an average, one obtains the
following results:
\be
<t_1^{\mu\nu}> = - 2k^\mu k^\nu e^{\ld\rho} e^*_{\ld\rho} + \frac{1}{2}k^\mu k^\nu 
e^\ld_\ld e^{*\a}_\a.
\ee
$$
<t_2^{\mu\nu}> = \frac{1}{2} k^\mu k^\nu e^\ld_\ld e^{*\a}_\a.
$$

Suppose one observes    this radiation in the wave zone, one can write  
the polarization tensor in terms of the Fourier transform of $T_{\mu\nu}$:
\be
e_{\mu\nu}({\bf x},\omega)=\frac{g}{4\pi r}[T_{\mu\nu}({\bf k},\omega) - 
\frac{1}{2} \eta_{\mu\nu}T ({\bf k},\omega)], \ \ \ \  T= T^\ld_\ld,
\ee
\be
T_{\mu\nu}({\bf k},\omega) \equiv \int d^3{\bf x}' T_{\mu\nu}({\bf 
x}',\omega)]exp(-i{\bf k} \cdot{\bf x}'), 
\ee
\be   \phi_{\mu\nu} ({\bf 
x,t}) \approx e_{\mu\nu}({\bf x}, \omega) exp(-ik_\ld x^\ld) + c.c.,
\ee
where we have used (13) and (14) with the approximation $|{\bf x} - {\bf 
x}'| \approx r -{\bf x}' \cdot {\bf x}/|{\bf x}|$ and ${\bf k} =\omega {\bf 
x}/|{\bf x}|$ in the wave zone.
Thus, the average energy-momentum of a gravitational plane wave can be 
written as
\be
<t^{\mu\nu}> = -\frac{G_{N}}{\pi r^2}k^\mu k^\nu \left(
T^{\ld\rho}({\bf k},\omega)T^*_{\ld\rho}({\bf k},\omega)
 - \frac{1}{2}T({\bf k},\omega)T^{*}({\bf k},\omega) \right).
\ee

The power $P_o$ emitted per unit solid angle in the direction ${\bf x}/|{\bf x}|$ 
is~\cite{8}
\be
\frac{dP_o}{d\Omega}= r^2 \frac{x^i <t_{i0}>}{|{\bf 
x}|}.
\ee
It can be written in terms of $T({\bf k}, \omega)$ in (23),
\be
\frac{dP_o}{d\Omega}= \frac{G_{N} \omega^2}{\pi}\left(T^{\ld\rho}({\bf k},\omega)
 T^*_{\ld\rho}({\bf k},\omega) -
 \frac{1}{2}T({\bf k},\omega)T^{*}({\bf k},\omega) \right).
\ee
Although the energy-momentum tensor of the gravitational field (18) 
in Yang-Mills gravity is quite different from that in general 
relativity, the result (27) for the power emitted per solid 
angle turns out to be the same as that 
obtained in general relativity and consistent with 
experiments.\cite{8,9}  Following the usual method  and 
approximation, one can calculate the power radiated by a body rotating
around one of 
the principal axes of the ellipsoid of inertia.  At twice the
 rotating frequency $\Omega$,
 i.e., $\omega=2\Omega$, one obtains the total power $P_o(\omega)$
emitted by a rotating body:
\be
P_o(2\Omega) = \left[ \frac{32G_{N} \Omega^6 I^2 e_{q}^2}{5} 
\right], \ \ \ \ \ \   c=1.
\ee
where $I$ and $e_{q}$ are respectively moment of inertia and equatorial 
ellipticity.  Thus, to the second order approximation, the
 gravitational quadrupole radiation (28) 
predicted by the Yang-Mills gravity is also the same as 
that predicted by general relativity.\cite{8}  

\section{Accelerated Cosmic Expansion and the Effective Repulsive 
Force due to Cosmological Constant}

The discovery of the accelerated cosmic expansion
stimulated many discussions.\cite{10}    The physical 
origin of a new `repulsive force' (in the Newtonian approximation)
has not been established.   We would like to discuss and compare two 
of suggestions for the physical origin of the new repulsive force: (A) the 
cosmological constant in Einstein's  field equation, and (B) the 
Lee-Yang gauge field associated with the conserved baryon number.  In 
this section, we first review and discuss the effective repulsive 
force due to the cosmological constant from the viewpoint of field 
theory.

It is reasonable to expect that in a
non-relativistic approximation of a theory, apart from the usual
Newtonian gravitational force, there is an additional repulsive force
between two ordinary objects.  The total `cosmic force' $F_{C}$ between two
objects can be written phenomenologically as a combination
of the usual gravitational attractive force and another long-range 
repulsive force $Bf(r)$,
\be
F_{C} \approx  - \frac{G M_1 M_2}{r^2} + Bf(r),
\ee
where B denotes the strength of the new long-range force.    

Experimentally, this new force in (29) must be very much smaller than 
the Newtonian gravitational force in the solar system and in our galaxy because it has 
not been detected.  It appears that this new force becomes important 
only in a very large cosmic scale.

Although we do not assume general relativity or Einatein's equation 
in the formulation of Yang-Mills gravity, it is interesting to 
compare it and the corresponding results in the present theory. Einstein's 
field equation with the cosmological constant is given by
\be
R^{\mu\nu} -\frac{1}{2}g^{\mu\nu}R - \ld g^{\mu\nu}= -8 \pi G T_{E}^{\mu\nu}.
\ee
In order to get a simple picture for the role played by the cosmological 
constant $\ld$, let us consider the static Newtonian approximation 
of (30):\cite{11}
\be
\bigtriangledown^2 \phi = m \delta^{3}({\bf r}) + \ld,
\ee
for a mass point located at the origin.  We have used $T_{E}^{00}=m\delta^{3}({\bf 
r})$ and $g^{00}=1 - 2\phi$. 
The spherically symmetric solution to (31) is given by
\be
\phi= \phi_{g}+\phi_{c},  \ \ \  \phi_{g}=- \frac{Gm}{r},  \ \ \ \  
\phi_{c} = \frac{\ld}{6} r^{2},
\ee
 where $\ld < 0$ corresponds to a cosmic repulsive force in a 
 classical limit.
The cosmological 
constant $\ld$ in (31) behaves like an undetectable `new ether' with a  constant density 
everywhere in the universe.  
It acts like a strange source 
which generates a new cosmological potential $\phi_{c}= \ld r^{2}/6$.
 
As  a result, the equation of motion of a freely moving test particle in the 
Newtonian limit is changed to the following form
\be
\frac{d^{2} {\bf r}}{dt^{2}}={\bf g} - C{\bf r},
\ee
where ${\bf g}$ is the gravitational acceleration produced by the 
distribution of ordinary matter, while the `dark energy' acceleration 
$\propto {\bf r}$ is due to a constant `dark energy' density 
$\Omega_{\Ld o}$ everywhere in the universe ($C=\Omega_{\Ld o}H^{2}_{o}$ and 
$H_{o}$=Hubble constant).\cite{10}   

Einstein originally believed (in 1917) that the large-scale structure of the 
universe is static, so he introduced the term with the cosmological 
constant in his field equation (30) to be consistent with his belief.  However, 
this static solution for a universe is unstable.  This property can 
be seen in equation (33), 
the mass distribution can be chosen such that the two terms cancel so 
that ${d^{2} {\bf r}}/{dt^{2}}=0$.  However, the cancellation of 
these two terms in (33) can be easily upset by a redistribution of 
masses in the universe or by a small perturbation to the mean mass 
density.\cite{10}  Eventually, Einstein gave up the cosmological constant for two 
reasons:(i) logical economy, and (ii) Hubble's discovery of the 
expansion of the universe.

Nevertheless, the presence of the cosmological   constant $\ld$ in 
Einstein's field equation (30) is now postulated by many people to be the cause of the 
observed accelerated expansion of the universe.  Although Einstein 
did not consider  the cosmological constant to be part of the 
energy-momentum tensor, it is equivalent to consider it as part of 
the energy-momentum tensor, i.e., a `new component' in the content of 
the universe.\cite{10}  But from the viewpoint of   field theory, the presence 
of the `additional' source term $\ld$ in (31) turns out to be 
very strange.  Equation (31) suggests that the 
potential field $\phi$ (or $g_{00}=1 + 2\phi$) is generated by two 
distinct sources, $m \delta^{3}({\bf r})$ and $\ld$, at 
the same time.  Nevertheless, 
so far all known experiments show that different kinds of sources generate different 
kinds of fields in field theory and particle physics.  Furthermore, this type of 
non-local source will probably 
cause further difficulty in the quantization of field in (30).  For 
example, the inverse-square force corresponding to the potential ${Gm}/{r}$ 
has the  field-theoretic interpretation, namely, it is due to 
the exchange of virtual 
gravitons described by the field equation (30), which can be seen in 
Feynman's discussion.\cite{12}  On the contrary,  
the linear   force in (33) corresponds to the cosmological potential 
$r^{2} \ld/6$ and does not have a field-theoretic interpretation based on 
the field equation (30).   

\section{Accelerated Cosmic Expansion and the Lee-Yang Force Associated 
with Conserved Baryonic Charge}

The theory of Yang-Mills gravity is based on the translation gauge group 
$T(4)$ in flat spacetime and does not allow a cosmological constant. 
The motivation to find  a more natural explanation for accelerated 
cosmic expansion led us to 
investigate  the cosmic Lee-Yang repulsive force associated with the 
observed conservation of baryon numbers (or baryon charges)
through the principle of gauge symmetry.\cite{13}  For 
this purpose, the Yang-Mills gravity is extended to include the 
$U_{b}(1)$ gauge field, so that the whole theory is based on the gauge group
$T(4)\times U_{b}(1)$.

Soon after the creation of Yang-Mills theory of SU(2) gauge field related to 
isospin conservation, Lee and Yang discussed in 1955 a 
long-range repulsive force ($\propto 1/r^{2}$) between 
baryons based on the $U_{b}(1)$ gauge symmetry associate
with the experimentally established conservation of baryon charge (or 
number).\cite{14}  Using E\"otv\"os 
experiment, the strength of such a repulsive force between nucleons 
(or baryons) was estimated
to be at least one million times smaller than that of the gravitational force.  
Such an extremely weak inverse-square force will probably never be observed.  
Nevertheless, we discuss a modified gauge invariant Lagrangian    
for the $U_{b}(1)$ gauge field, which
suggests  a new r-independent
cosmological force $Bf(r)$ in (29) 
between observable galaxies (which are assumed to be made of baryons 
and leptons).  We suggest that such a new r-independent cosmic force can  
be produced by the gauge  fields
  associated with baryon numbers
and electron-lepton numbers.\cite{13}  The conservation of these 
quantum numbers has been experimentally established.\cite{15}   

We observe that the gauge invariant Lagrangian for massless
$U_{b}(1)$ gauge field is, strictly 
speaking, not unique.  The reason is that  
besides the usual Lagrangian which is quadratic in the
fields strength $B_{\mu\nu}$, there is another simple gauge invariant
  Lagrangian
which is quadratic in $\p_\ld B_{\mu\nu}$.\cite{13}  
This simple  gauge invariant Lagrangian is interesting because it
can lead to a `linear
  potential', $\propto r$, which differs from the `quadratic 
  potential' in (32) associated with  the 
cosmological constant.  

Let us consider such a gauge invariant Lagrangian involving up and 
down quarks and baryonic gauge field $B_{\mu}$ for accelerated 
cosmic expansion (ACE):
\be
L_{ACE} = -\frac{L_s^2}{4} \p_\ld B_{\mu\nu} \p^\ld B^{\mu\nu} 
 + L_{ud} ,
\ee
$$L_{ud}=i\overline{u}_{n} \g_\mu 
(\p^\mu-\frac{ig_{b}}{3}B_{\mu})u_{n} - 
m_{u}\overline{u}_{n}u_{n} +
i\overline{d}_{n} \g_\mu (\p^\mu-\frac{ig_{b}}{3}B_{\mu})d - 
m_{d}\overline{d}_{n}d_{n},$$
 where the color index $n$ is summed 
from 1 to 3.  This is the part of the Lagrangian (1) with $U_{b}(1)$ 
gauge symmetry.  As usual, it is not necessary to symmetrize the 
fermion Lagrangian with the $U_{b}(1)$ gauge symmetry.

One can easily include the gauge field associated with 
the conserved lepton numbers in (34).  For simplicity, we shall not 
discuss it here.  The new gauge-invariant field equation derived from 
(34) is a  
fourth-order partial differential equation of $B_{\mu}$,
\be
\p^2\p_\nu B^{\nu\mu} - g'_b J_{q}^\mu = 0, \ \ \ \ \   g'_b = g_b/(3 L_s^2),
\ee
where the source of the gauge field is 
$$J_{q}^{\mu} =  \overline{u}_n \g^{\mu}u_n  +
  \overline{d}_n \g^{\mu}d_n.$$
Clearly, the field strength $B_{\mu\nu}$  satisfies the Bianchi identity
\be
\p^\ld B^{\mu\nu} +\p^\mu B^{\nu\ld} +\p^\nu B^{\ld\mu} = 0.
\ee
For the static case, the field $B_{0}$ satisfies the fourth order 
differential equation
\be
\bigtriangledown^2 \bigtriangledown^2 B^0 
= -\frac{g_{b}}{3L_{s}^{2}}J^{0} \equiv \rho_{B}.
\ee
The general solution to $B^{0}$ can be expressed in terms of a 
modified version of the usual Green's function for a second order 
partial differential equation,
\be
B^{0}({\bf r}) = \int F({\bf r, r'}) \rho_{B}({\bf r'}) d^{3}r',
\ee
$$F({\bf r, r'}) = \frac{1}{\bigtriangledown^2}\frac{-1}{4\pi |{\bf 
r-r'}|} =\int \frac{e^{-i{\bf k} \cdot ({\bf r - r''})}}{{\bf 
k^{2}}}\frac{d^{3} k}{(2\pi)^{3}}\frac{1}{ 4\pi |{\bf r''-r'}|} d^{3}r'',$$
where $1/{\bigtriangledown^2}$ is an integral operator.

Suppose we impose a `Coulomb-like gauge' $\p_k  B^k = 0$, the static exterior potential
satisfies the equation $\bigtriangledown^2 \bigtriangledown^2 B^0 = 
0$. The solution for such a potential 
can be written in the form $B^{0}=A'/r + B'r + C'r^2$.
Different $r$-dependent terms in $B^{0}$ corresponds to sources with different
types of singularity at $r=0$ or different boundary conditions at
infinity.  For large distances, the last term $C'r^2$ dominates.  In 
this case, the classical potential
$B_0$ in (37) leads to the same effect as that of
Einstein's cosmological constant and implies 
the modified gravitational law of motion (33).  However, the potential
$C'r^2$ is incompatible with the field equation (35) [with the usual 
current-source $J^{\mu}_{q}$] from the 
viewpoint of field theory: 
There is simply no way to produce such a force through the exchange 
of virtual quantum between two Dirac's fermions carrying baryon 
charges.  It turns out that only the linear potential 
$B'r$ is due to the exchange of virtual quantum described by 
the fourth order field equation (37).  Why?  The reason is as follows:

From the viewpoint of quantum field theory,
the fermion (baryon) source $\rho_{B}$ in (37) is represented by the usual 
delta-function because these fermions satisfy the Dirac 
equations.  Based on this property, we can show that
the solution for the potential $B_0$ should be proportional to $r$
rather than $r^2$.   This result can be seen explicitly by
substituting $\rho_{B}({\bf r})=g'_{b}\delta^{3}({\bf r})$ in (38), 
where $g'=g_{b}/(3L_{s}^{2})$ for quarks and $g'=g_{b}/(L_{s}^{2})$ for 
protons and neutrons.   We obtain
\be
B^{0}({\bf r}) =g'_{b}\int_{-\infty}^{\infty}{\frac{1}{({\bf k}^2)^2} e^{i{\bf k}\cdot{\bf r}} d^3
k} = - \frac{g'_{b}}{8\pi}|{\bf r}|,
\ee
which can be understood as a generalized function.\cite{16}  It 
can also be obtained by using the relation
\be
\int_{0}^{\infty} \frac{sin ax}{x(b^{2}-x^{2})}dx 
=\frac{\pi}{2b^{2}}(1-cos ab),
\ee
and take the limit $b \to 0$.  The limit appears to be not trivial 
because if the denominator $x(b^{2}-x^{2})$ in the integrand of (40) is 
replaced by $(b^{4}+x^{4})/x$, the integral exists, but the limit
$b \to 0$ does not.  Thus, it is more rigorous to treat the Fourier transform of 
linear potential and its inverse transform within the 
framework of generalized functions.\cite{16}

The result (39) implies that the force associated with the linear 
potential $B^{0}({\bf r})$ can be interpreted as the exchange of 
virtual quantum between two quarks, where the quantum satisfies the 
fourth-order field equation (37).  The situation is the same as that in quantum 
electrodynamics.  Namely, the
static Coulomb potential produced by a point charge is
the Fourier transform of the Feynman
propagator of a virtual photon with $k_{0}=0$. 

\section{Experiments of Lee-Yang Force with the Wu-Doppler Effect}

Let us consider a possible experimental tests of Lee-Yang force by measuring the 
accelerations of supernovae based on a modified Doppler Effect.  First we 
note that, in analogy with (33), the constant Lee-Yang force 
will modify Newtonian law of motion for a 
test particle as follows:
\be
\frac{d^2{\bf r}}{dt^2} = {\bf g} +  {\bf g}_u,
\ee
It is convenient to use a nucleon rather than a quark with a baryonic 
charge $g_{b}/3$ for discussions.  A nucleon carries a baryon charge 
$g_{b}$  and a mass m.  Thus, the second term for two nucleons will be
$|{\bf g}_u| =  g_b/(8\pi L_s^2 m)$, which
is  the r-independent acceleration associated the cosmic linear 
potential
$g_{b}^{2}{\bf r}/(8\pi L_{s}^{2})$ of the fourth-order
field equation (37) [with $g_{b}/3$ replaced by $g_{b}$ of a baryon]. 
It implies the existence of a
 constant acceleration ${\bf g}_u$ between any two galaxies made of 
baryons, provided other
forces (e.g., the `magnetic-type' force) are negligible.  This is an 
interesting
prediction of $U_{b}(1)$ gauge symmetry together with the fourth-order
gauge-invariant field equation (37).  It is important  to test 
the prediction (41).  
 
Physically, the accelerated cosmic expansion\cite{17} implies that, 
strictly speaking, 
there is no inertial frame in the universe.  Inertial frames are 
merely 
idealized reference frames  for simplification of discussions of the 
physical laws and phenomena.  At least, one should
seriously consider physics in a more realistic reference frames 
with linear accelerations by investigating a generalization of
Lorentz transformations for accelerated frames.
We have considered one of simple generalizations, 
called Wu transformations, for reference frames with 
constant-linear-accelerations (CLA) in the previous paper.\cite{5}
In terms of the differentials 
$dx^{\mu}_{I}=(dw_{I}, dx_{I}, dy_{I}, dz_{I})$ for an `Inertial 
Frame' 
$F_{I}(x_{I}^{\mu})$ 
and $dx^{\mu}=(dw, dx, dy, dz)$ for a CLA 
frame $F(x^{\mu})$, a simple generalization of the Lorentz 
transformation takes the following form:
\be
dw_I = \g (Wdw + \b dx) , \ \    dx_I = \g(dx + \b W dw) \\ 
\ee
$$dy_I  = dy ,  \ \ \ \ \   dz_I  = dz ; $$  
\noindent
where $W=\g^2(\g_o^{-2} + \a_o x)$,	$\b = \a_o w + \b_o, \ \  	\g_o =
1/\sqrt{1-\b_o^2}  ,  \ \ 	\g =1/\sqrt{1-\b^2}.$ They can be 
integrated to obtain the transformation for $x^{\mu}_{I}$ and 
$x^{\mu}$.\cite{5}  The Wu transformation (42) implies
\be
ds^{2}=dw_{I}^{2} - dx_{I}^{2} - dx_{I}^{2} - dx_{I}^{2}
=W^{2}dw^{2}- dx^{2}- dy^{2}- dz^{2}=P_{\mu\nu}dx^{\mu}dx^{\nu}.
\ee
where $P_{\mu\nu}=(W^{2},-1,-1,-1)$ is the metric tensor for the 
CLA frame $F$.  This simple form of the metric tensor $P_{\mu\nu}$ 
suggests that if one defines a distorted `Wu differential', (Wdw, 
dx, dy, dz), for the CLA frame $F(x^{\mu})$, then the transformation of the 
`Wu differential' from 
such a CLA frame to an inertial frame $F_{I}$ will be formally the 
same as the Lorentz transformation, except that the constant velocity is 
replaced by a time-dependent velocity $\b = \a_o w + \b_o$, as shown 
in (42).  Nevertheless, the accelerated frame $F(w,x,y,z)$ is not 
equivalent to the inertial frame $F_{I}(w_{I},x_{I},y_{I},z_{I})$ 
because the Wu differentials are distorted only for CLA frames due to 
its acceleration, but not 
for inertial frames.

However, it appears reasonable to assume that all reference frames with 
the same constant-linear-acceleration (CLA) are `equivalent' in the 
following sense:   Suppose $F$ and 
$F'$ are two such CLA frames.  An atom $H_{e}$ at rest in $F$ has the 
same properties as another $H_{e}$ atom at rest in $F'$.  For 
example, if an atom $H_{e}$ at rest in $F$ ($F'$) emits a light wave with
the wavelength $\ld_{o}$ ($\ld'_{o}$), as immediately measured in $F$ ( $F'$); 
then we have the equality relation, $\ld_{o} = \ld'_{o}$.  It appears 
to be reasonable to employ such an 
`equivalence' for two CLA frames as a guiding principle 
for our discussions below. 

To measure directly the accelerations of distant supernovae Ia, the usual 
 Doppler effect is no longer adequate.  One must know how the 
usual relation of the Doppler effect is modified by the linear 
accelerations of the light source and observers. In this 
connection, the accelerated Wu transformation will be very useful. 
Suppose a light source is at rest in the 
accelerated frame $F$, and the observer is, for simplicity, at rest 
in an inertial frame $F_{I}$.
The covariant wave 4-vector $k_{\mu}$ has the same transformation as the covariant 
coordinate 4-vector $dx_{\nu}=P_{\nu\mu}dx^{\mu}$.  We note that the coordinate 
$x^{\mu}$ of a CLA frame is no longer a 4-vector.  Thus the Wu 
transformation implies the following accelerated Wu-Doppler effect:
\be
k_{I0} = \g (W^{-1}k_{0} - \b k_{1}) , \ \   
k_{I1} = \g(k_{1} - \b W^{-1} k_{0}) \\ 
\ee
$$k_{I2}  = k_{2} ,  \ \ \ \ \   k_{I3}  = k_{3} ; $$  
for the covariant wave 4-vector $k_{\mu}=(k_{0}, k_{1},k_{2},k_{3})$, 
where 
\be
(W^{-1}k_{0})^{2} -{\bf k}^{2} = k_{I0}^{2} - {\bf k}_{I}^{2}.
\ee
In the limit of zero acceleration, $\a_o \to 0$, the Wu 
transformation reduces to the Lorentz transformation,  and the 
Wu-Doppler effect (44) becomes the usual relativistic Doppler effect in 
special relativity.

 Presumably, the Lee-Yang force and the constant acceleration $g_{u}$ 
 in (41) due to the baryonic charge are
extremely small, so that their dynamical effects in particle physics 
cannot be detected in high energy laboratories.   Moreover, only  
gigantic bodies like galaxies separated by a great distance can 
have enough repulsive Lee-Yang force to 
overcome the gravitational attractive force and, hence, to move with an
acceleration which may be detected through the Wu-Doppler effect of the 
wavelength in the radiation of a supernova Ia.

Suppose the earth, the supernova `a' and the supernova `b' are 
respectively at rest in the constant-linear-acceleration (CLA) 
frames $F$, $F'$ and $F''$, which are moving with velocities 
$\b=\b_{o}+\a_{o} w$, $\b'=\b'_{o}+\a'_{o} w'$, and 
$\b''=\b''_{o}+\a''_{o} w''$ along the +x axis.  The Wu-Doppler effect 
(44) can be applied to these  three CLA frames.  Suppose an atom at 
rest in $F$ ($F'$, $F''$) emits  a lights (propagating 
along the x-axis) with the wavelength $\ld_{o}$ ( $\ld'_{oa}$, $\ld''_{ob}$), as 
immediately measured in $F$ ($F'$, $F''$).  We have 
$\ld_{o}=\ld'_{oa} = \ld''_{ob}$ if  the 
frames $F$, $F'$ and $F''$ have the 
same acceleration or have very small accelerations.  
Suppose   these two different lights emitted from supernovae have the 
wavelength $\ld_{a}$ and ($\ld_{b}$), as measured on the earth (i.e., 
$F$ frame).  We obtain
\be
\frac{\ld_{a}-\ld_{b}}{\ld_{a}\ld_{b}} = \frac{1}{\ld_{o}\g(1-\b)} 
[\g''(1-\b'')- \g'(1-\b')]
\ee
from (44) and (45).  In general, the values of accelerations of the 
supernovae are extremely small and difficult to measure because the 
initial conditions of 
their motions in (46) are not known.\cite{18}  However, we are 
interested in testing the predictions (33) and (41), namely, whether 
the accelerations $\a'_{o}$ and $a''_{o}$ are the same or not.  To 
see the difference of these two cases based on the Wu-Doppler 
effect (46), we assume that, for simplicity, the Earth 
frame can be approximated by an 
inertial frame, $F=F_{I}$, and that the initial velocities of the 
frames $F'$ and $F''$ are zero. Using the inverse Wu transformations, 
the velocities $\b'$ and $\b''$ can 
be expressed in terms of quantities measured in the frame 
$F=F_{I}$, 
e.g., $\b'=(\a_{oa}w_{Ia}+\b_{o}/\g_{oa})/(\a_{oa}x_{Ia}+1/\g_{oa})$.\cite{5,19} 
For small velocities and accelerations, (46) can be approximated by
\be
\frac{\ld_{a}-\ld_{b}}{\ld_{a}\ld_{b}} = \frac{1}{\ld_{o}}
\left[\frac{\a_{oa}w_{Io}}{1 + \a_{oa}x_{Ia}} - 
\frac{\a_{ob}w_{Io}}{1 + \a_{ob}x_{Ib}}\right],
\ee
where $w_{Io}$ is the time of observation, and $x_{Ia}$ and $x_{Ib}$ 
are the distances of the supernovae `a' and `b' as measured in the Earth frame.
If supernovae `a' and `b' have the same acceleration, 
$\a_{oa}=\a_{ob}=g_{u}$, the value of cosmic 
acceleration can be estimated through 
the measurement of $\ld_{a}$ and $\ld_{b}$ on the Earth, provided 
(A) the Earth can be considered as the $F_{I}$ frame after some necessary 
and careful corrections and (B) $w_{Io}$ is measured in cosmic time 
which is presumably the same order of magnitude as the age of the 
universe.  

When one considers the motion of galaxies, the expansion and the accelerated 
expansion of the universe as a whole, one uses a cosmic time.  Such a cosmic 
time appears to be quite different from the relativistic time of 
special relativity.  
However, within the four-dimensional symmetry framework, 
one can define a time for all observers in different frames to record 
time and to describe 
physics, and still preserve the Lorentz and Poincar\'e invariance of 
physics laws.  Such a 
time is called `common time' which resembles the cosmic 
time.\cite{19,20}

On the 
other hand, if the accelerations $\a_{o}$, $\a'_{o}$, and $\a''_{o}$ are 
not the same, it will be more difficult to determine the values of 
accelerations by the Wu-Doppler effect.  Nevertheless, since all 
these accelerations are very small, (47) may be approximated by
\be
\frac{\ld_{a}-\ld_{b}}{\ld_{a}\ld_{b}} = \frac{w_{Io}}{\ld_{o}}
\left[\a_{oa}-\a_{ob} +\a_{ob}^{2}x_{Ib}-\a_{oa}^{2}x_{Ia} \right].
\ee
The 
qualitative difference between (33) and (41) can be tested by using 
(48), provided 
one has enough data and accuracy for the measurements of the wave 
lengths.   Of course, before one applies the Wu-Doppler 
effect (44) to investigate the acceleration of cosmic expansion, one must 
also test (44) in the laboratory.  

It is important to carry out many different kinds of
experiments to test the difference between two interesting  
predictions (33) and (41) for the accelerated cosmic expansion.   
Qualitatively speaking, the universe was much smaller in an earlier 
era, so that the Lee-Yang repulsive force between two galaxies was overwhelmed 
by the usual attractive gravitational force.  Thus we have 
decelerated cosmic expansion in an earlier era.  As the intergalactic 
distances increase, the gravitational force decreases, and
there will be a critical distance $R_{c}$ where the Lee-Yang repulsive 
force cancels the gravitational attractive force between two galaxies.  
As an example, if one considers 
an isolated system of 
two baryons with baryon charge $g_{b}$ and mass m, the critical 
distance $R_{c}$ is given by
\be
R_{c} = (Gm^{2} 8\pi L_{s}^{2}/g_{b}^{2})^{1/2}.
\ee
When the intergalactic distances became larger  than the critical 
distance, the r-independent Lee-Yang repulsive force would 
overcome the gravitational attractive force,
 and one would have accelerated cosmic expansion. This appears to be 
what we have observed in the recent era of cosmic evolution.\cite{17}

Comparing the difference of the two predictions (33) and (41), 
we could have the following scenario:
As the intergalactic distance increases, the constant 
Lee-Yang force may be weaker than the linear force.  One usually takes the data of 
distant supernovae Ia (with roughly the same intrinsic brightness) and plots relative 
apparent brightness (or relative brightness) against the redshift z to show the accelerated 
cosmic expansion at the present era. But the data also show that in an earlier 
epoch (in cosmic time) with a higher redshift z, one has deceleration.\cite{17}   
In comparison with the linear force by using the graph 
of relative apparent brightness and redshift, 
the constant Lee-Yang force will probably  give a smaller 
acceleration for the cosmic expansion for small redshift and give a larger deceleration
for large redshift z (or earlier era).  We stress that the experimental 
evidence for unambiguous r-dependence or r-independence of the cosmic acceleration
is important because of the 
following reasons: If the prediction (33) of the cosmological constant
is confirmed, then the current field 
theory and particle physics are inadequate for understanding physics 
at the cosmological scale.  On the other hand, if the prediction (41) 
of the Lee-Yang force associated with conserved baryonic charge is confirmed, 
this result would imply that we do not have to make the unnatural 
assumption that about 70\% of 
the energy density of the universe is the unknown `Dark 
Energy'\cite{17,21} or a sort of `new 
ether' everywhere in space.  Furthermore, this would  imply that we 
can understand cosmological phenomena based on  
field theory and particle physics, which are originally formulated for 
microscopic world. This would be a further support of the principle 
of gauge symmetry and an interesting unification of 
physics at microscopic and macro-cosmic worlds.

\section{Discussions and Remarks}

(A) Effective Metric Tensor

The basic Lagrangian for tensor fields and electromagnetic 
fields\cite{5} 
within Yang-Mills gravity does not explicitly involve the effective 
Riemannian metric 
tensor.  From the viewpoint of gauge symmetry, the Yang-Mills
gravity in flat spacetime reveals the 
field-theoretic origin of an effective Riemannian 
metric tensor for the motion of classical particles.  
Namely, such an effective metric tensor  
shows up only in the limit of geometric optics (or classical limit) 
of wave equations for quantum particles.  In other words, the
Yang-Mills approach to 
gravity suggests that the underlying basis for gravity is the 
translation gauge symmetry in flat spacetime rather than the general 
coordinate invariance in curved spacetime.  This property could shed 
light on quantum gravity, which will be discussed in a separate paper.

(B) Fundamental Length

The gauge group of the Yang-Mills gravity is $T(4)\times U_{b}(1)$, 
where $T(4)$ and $U_{b}(1)$ appear to be simply juxtaposed, in 
contrast to those in the unified electroweak theory.  
However, there are two basic constants with the dimension of length 
in this theory:
One is the gravitational coupling constant $g$ in the $T(4)$-invariant Lagrangian, 
and the other is the length scale $L_{s}$ in the $U_{b}(1)$-invariant 
Lagrangian.  It is quite possible that these two basic constants are 
not independent.  And they could provide a profound relationship 
that reveals the fundamental length of nature.

(C) Linear Potentials for Accelerated Cosmic Expansion and for Quark 
Confinement

The physical significances of linear potential in field theory are [I] 
it can provide a constant repulsive force for the accelerated cosmic expansion 
(at extremely large distances), and [II] it can also provide an 
understanding of the permanent confinement of quarks and antiquarks inside hadrons 
(at very small distances), similar to Yukawa's treatment for the 
short-range nuclear force.  The property [II] is not trivial because if 
the gauge field is a vector field, then the linear potential will 
have both attractive and repulsive force.  But the quark confinement 
requires the forces for quark-quark and quark-antiquark to be 
always attractive.  One simple way to satisfy this requirement 
appears to be 
that the field is, say, a scalar  field\cite{22} or a tensor field, 
similar to the spacetime  translation gauge field.\cite{5}
It would be very interesting and significant if the potentials at 
the smallest distance scale between confined quarks~\cite{23} and at
the largest scale between galaxies with accelerated 
expansion are both due to linear potentials associated with the 
fourth-order field equations.  These properties deserve to be further studied.
 
\bigskip

{\bf Acknowledgements}
\bigskip

The author would like to thank D. Fine and L. Hsu for useful 
discussions.  The work was supported in part by the Jing Shin
Research Fund of the UMass Dartmouth Foundation.


\bibliographystyle{unsrt}

\end{document}